# Quantum hydrodynamics of electron gases

Radomir Slavchov and Roumen Tsekov
Department of Physical Chemistry, University of Sofia, 1164 Sofia, Bulgaria

Electron gases in metals are described as quantum charged Newtonian viscous fluids experiencing Ohmic Darcy friction on the solid lattice ions as well. The dispersion relation of the electron acoustic waves is derived, which shows the existence of new quantum diffusion processes. The electric double layer near a metal surface is studied, which exhibits a quantum oscillatory-decaying behavior different from the Friedel oscillations.

Quantum fluids attracted scientific interest due to important quantum phenomena such as superfluidity of liquid helium and superconductivity of electrons in metals and semiconductors. Traditionally they are described via nonlinear Schrödinger equations [1, 2]. In the present paper an alternative approach to electron gases is employed, which is based on quantum hydrodynamics [3-9]. Additionally to electrostatics, the electron-phonon interaction is modeled as an Ohmic Darcy friction, while the electron-electron collisions are accounted for via two specific viscosities of the electron gas [10-12]. The present analysis is comparable to quantum magneto-hydrodynamics [13, 14] at negligibly low magnetic fields. The dispersion relation of the acoustic waves in electron gases of metals is obtained, which shows the existence of new kinds of quantum diffusion processes. In general, a more rigorous description can be achieved by the Wigner-Poisson equation, which is, however, difficult to solve. Derivation from it of the quantum hydrodynamics shows that higher-order hydrodynamic terms appear as well [8, 15]. This is, however, the case also in classical hydrodynamics, which describes well systems not very far from equilibrium.

In the frames of the Navier-Stokes electro-hydrodynamics [15] the motion of an isothermal charged viscous fluid is governed by the following continuity and dynamic equations

$$\partial_t \rho = -\nabla \cdot (\rho V) \tag{1}$$

$$\partial_t V + V \cdot \nabla V = -\nabla \mu / m + \nu \nabla^2 V + (\xi + \nu/3)\nabla \nabla \cdot V \tag{2}$$

$$\varepsilon_0 \varepsilon \nabla^2 \phi = e(\rho - \bar{\rho}) \tag{3}$$

Here $\rho$ is the local fluid density, $V$ is the hydrodynamic velocity, $m$ and $e$ are the unit mass and charge, $\mu$ is the electro-chemical potential, and $\nu$ and $\xi$ are the kinematic shear and dilatational viscosities, respectively. In the electrostatic Poisson equation (3) $\phi$ is the electric po-tential and $\bar{\rho}$ is the average electron density being equal in the frames of the jellium model also to the mean density of positive charges originating from the lattice ions. To close this system of five equations an equation of state of the electron gas is required in the form $\mu(\rho,\phi)$. In a previous paper [16] the electro-chemical potential of a non-uniform electron gas is derived from the Thomas-Fermi-Dirac-Weizsäcker density functional theory [17]. In the case of relatively dilute electron gases it reduces to the expression

$$\mu = \hbar^2 (3\pi^2 \rho)^{2/3} / 2m + Q + k_B T \ln \rho - e\phi \qquad (4)$$

where the first term is the local Fermi energy, while $Q \equiv -\hbar^2 \nabla^2 \sqrt{\rho} / 2m\sqrt{\rho}$ is the Bohm quantum potential originating from the Weizsäcker correction or Fisher entropy. Due to considered low electron density the Dirac exchange energy contribution to the electro-chemical potential $\mu$ is neglected. Introducing Eq. (4) in Eq. (2) the latter changes to

$$\partial_t V + V \cdot \nabla V + \gamma V = -\nabla [\hbar^2 (3\pi^2 \rho)^{2/3} / 2m + Q + k_B T \ln \rho - e\phi]/m + \nu \nabla^2 V + (\xi + \nu/3)\nabla \nabla \cdot V \qquad (5)$$

Note that Eq. (5) is accomplished additionally by the linear term on $V$ accounting for the Ohmic friction of the electrons on the lattice ions, where $\gamma$ is a specific friction constant [9]. It is like the Darcy law in hydrodynamics since the electrons are moving in a porous media generated by the lattice ions. Thus Eq. (5) describes the irreversible dynamics of an electron gas due to dissipation of energy among the electrons and between the electrons and lattice ions vibrations.

In the case of small acoustic perturbations the local density, hydrodynamic velocity and electric potential can be written as proportional to harmonic perturbation amplitude $A$

$$A \exp(-i\omega t - iqx) = 1 - \rho/\bar{\rho} = (q/\omega)V = (\varepsilon_0 \varepsilon q^2 / e\bar{\rho})\phi \qquad (6)$$

The coefficients in Eq. (6) are chosen in such a way that $\rho$, $V$ and $\phi$ obey the linearized continuity (1) and Poisson (3) equations. Introducing these expressions in Eq. (5) and linearizing the result on the small amplitude $A$ yields the dispersion relation of the electron gas acoustic waves

$$\omega^2 + i\omega[\bar{\gamma} + (4\bar{\nu}/3 + \bar{\xi})q^2] = e^2 \bar{\rho}/\varepsilon_0 \varepsilon m + q^2 [\hbar^2 (3\pi^2 \bar{\rho})^{2/3} / 3m + k_B T]/m + q^4 (\hbar/2m)^2 \qquad (7)$$

where $\bar{\gamma}$, $\bar{\nu}$ and $\bar{\xi}$ are the average kinetic coefficients. Equation (7) provides several known limiting cases [2]. If the friction is negligible, this dispersion relation reduces to a well-known result [18]: it tends to the Langmuir frequency $\omega = \sqrt{e^2\bar{\rho}/\varepsilon_0\varepsilon m}$ for small wave numbers, to an acoustic frequency $\omega = q\sqrt{\hbar^2(3\pi^2\bar{\rho})^{2/3}/3m^2 + k_B T/m}$ with Bogolyubov thermo-quantum sound velocity for moderate wave numbers, while for large wave numbers Eq. (7) acquires the form corresponding to free electrons, $\hbar\omega = (\hbar q)^2/2m$. More interesting is the case of the high friction limit where the first inertial term in Eq. (7) can be neglected. In this case for moderate wave numbers Eq. (7) provides the diffusion law $i\omega = (D_Q + D)q^2$ with the quantum Bogolyubov $D_Q = \hbar^2(3\pi^2\bar{\rho})^{2/3}/3m^2\bar{\gamma}$ and classical Einstein $D = k_B T/m\bar{\gamma}$ diffusion constants, respectively. For large wave numbers another diffusion process appears with a quantum self-diffusion constant given by

$$D_Q^S = (\hbar/2m)^2/(4\bar{\nu}/3 + \bar{\xi}) \tag{8}$$

Since the diffusion coefficient and viscosity characterize the position and velocity dispersions of electrons, respectively, Eq. (8) could be interpreted as dissipative minimal Heisenberg relation [19].

Another interesting aspect in the present analysis is the equilibrium quantum hydrostatics [17]. Setting $\omega = 0$ the dispersion relation (7) reduces to

$$\lambda_Q^4 q^4 + (\lambda_{TF}^2 + \lambda_D^2)q^2 + 1 = 0 \tag{9}$$

where $\lambda_D = \sqrt{\varepsilon_0 \varepsilon k_B T/e^2\bar{\rho}}$, $\lambda_{TF} = \sqrt[6]{\pi^4\varepsilon_0^3\varepsilon^3\hbar^6/3m^3 e^6\bar{\rho}}$ and $\lambda_Q \equiv \sqrt[4]{\varepsilon_0\varepsilon\hbar^2/4me^2\bar{\rho}}$ are the Debye, Thomas-Fermi and a new quantum screening lengths, respectively. The latter is related to $\lambda_D$ and $\lambda_{TF}$ via the thermal de Broglie $\lambda_T = \hbar/2(mk_B T)^{1/2}$ and mean Fermi $\lambda_F = 2\pi/(3\pi^2\bar{\rho})^{1/3}$ wave lengths, i.e. $\lambda_Q^2 \sim \lambda_T\lambda_D \sim \lambda_F\lambda_{TF}$. The four solutions of Eq. (9) read

$$q\lambda_Q = \pm\sqrt{\pm\sqrt{\alpha^2 - 1} - \alpha} \tag{10}$$

where the dimensionless parameter $\alpha = (\lambda_{TF}^2 + \lambda_D^2)/2\lambda_Q^2$ represents a ratio between the screening lengths. Note that the physically relevant solutions from Eq. (10) are those with positive imaginary parts $q_{Im} > 0$ corresponding to decay of the electric potential far from the metal surface. In the case $\alpha < 1$ there are two proper solutions of Eq. (10) with the same positive imaginary part $q_{Im}\lambda_Q = \sqrt{(1+\alpha)/2}$ and opposite real parts $q_{Re}\lambda_Q = \pm\sqrt{(1-\alpha)/2}$. In this case the electric potential near the metal surface, being the solution of the linearized Poisson equation (3), exhibits an oscillatory-decaying behavior

$$\phi = A\sum_{k=1}^{2}(e\bar{\rho}/\varepsilon_0\varepsilon q_k^2)\exp(-iq_k x) = \phi_s \exp(-q_{\text{Im}}x)\cos(q_{\text{Re}}x) \tag{11}$$

Here $\phi_s = \phi(0)$ is the surface potential and its value depends on additional specifications of the metal interfacial properties. This oscillatory behavior is due to quantum effects, since in the classical limit the potential $\phi = \phi_s \exp(-x/\lambda_D)$ is purely decaying. Hence, it is distinct from the Friedel oscillations with a typical wave length $\lambda_F$. Mathematically the first quantum term in Eq. (9) is similar to the classical Cahn-Hilliard gradient term and for this reason the present results are comparable, for instance, to those in ionic liquids [20]. In the case $\alpha \geq 1$ the wave vector is purely imaginary $q\lambda_Q = i\sqrt{\alpha \pm \sqrt{\alpha^2-1}}$ and the electric potential exhibits simply exponential decay with two decay lengths. If $\alpha \gg 1$ these two characteristic screening parameters acquire the forms

$$q_{\text{Im}} = 1/\lambda_Q\sqrt{2\alpha} = 1/\sqrt{\lambda_{TF}^2 + \lambda_D^2} \qquad q_{\text{Im}} = \sqrt{2\alpha}/\lambda_Q = \sqrt{\lambda_{TF}^2 + \lambda_D^2}/\lambda_Q^2 \tag{12}$$

The first expression corresponds to a larger screening length, which in the classical limit reduces to the Debye length $\lambda_D$. The second short-length expression in Eq. (12) is essentially quantum and at zero temperature reduces to the Fermi wave vector.

[1]  J.F. Annett and H.H. Wills, *Superconductivity, Superfluids, and Condensates* (Oxford University Press, Oxford, 2004)
[2]  P.H. Roberts and N.G. Berloff, Lect. Notes Phys. **571**, 235 (2001)
[3]  F. Bloch, Z. Phys. **81**, 363 (1933)
[4]  R. Zwanzig, *Quantum Hydrodynamics*, Ph.D. Thesis (Caltech, Pasadena, 1952)
[5]  C.L. Gardner, SIAM J. Appl. Math. **54**, 409 (1994)
[6]  P. Degond and C. Ringhofer, J. Stat. Phys. **112**, 587 (2003)
[7]  R.E. Wyatt, *Quantum Dynamics with Trajectories: Introduction to Quantum Hydrodynamics* (Springer, New York, 2005)
[8]  K.H. Hughes, S.M. Parry and I. Burghardt, J. Chem. Phys. **130**, 054115 (2009)
[9]  A. Jüngel, *Transport Equations for Semiconductors* (Springer, Berlin, 2009)
[10] M.S. Steinberg, Phys. Rev. **109**, 1486 (1958)
[11] L.H. Hall, Phys. Rev. **153**, 779 (1967)
[12] R. Kishore and K.N. Pathak, Phys. Rev. **183**, 672 (1969)

[13]　F. Haas, Phys. Plasmas **12**, 062117 (2005)

[14]　B. Eliasson and P.K. Shukla, Plasma Fusion Res. **4**, 032 (2009)

[15]　I. Tokatly and O. Pankratov, Phys. Rev. B **60**, 15550 (1999); ibid. **62**, 2759 (2000)

[16]　R. Tsekov, Int. J. Theor. Phys. **48**, 2660 (2009)

[17]　E. Zaremba and H.C. Tso, Phys. Rev. B **49**, 8147 (1994)

[18]　G. Manfredi and F. Haas, Phys. Rev. B **64**, 075316 (2001)

[19]　S. Gao, Phys. Rev. Lett. **79**, 3101 (1997)

[20]　R. Tsekov, J. Chem. Phys. **126**, 191110 (2007)